\documentclass[aip,reprint,amsmath,amssymb]{revtex4-2}
\usepackage{subfigure}
\usepackage{epsfig}
\usepackage{epsf,epic}
\usepackage{color}
\usepackage{marvosym}
\usepackage{subfigure}
\usepackage{amsmath}
\usepackage{amssymb}
\usepackage{amsfonts}
\usepackage{wrapfig}
\usepackage{pstricks}
\usepackage{multirow}
\usepackage{bm}
\usepackage{dcolumn}
\newcommand{\etal}{\textit{et al.\ }}

\newcommand{\eg}{\textit{e.g.\ }}

\begin{document}
\title{N$_2$, NO and O$_2$ molecules in LiGaO$_2$ in both Ga and Li sites and
  their relation to the vacancies.}

\author{Klichchupong Dabsamut} 
\affiliation{Department of Physics, Faculty of Science, Kasetsart University, Bangkok 10900 Thailand}
\affiliation{Department of Physics, Case Western Reserve University, 10900 Euclid Avenue, Cleveland, Ohio 44106-7079, USA}
\author{Adisak Boonchun}
\email{adisak.bo@ku.th}
\affiliation{Department of Physics, Faculty of Science, Kasetsart University, Bangkok 10900 Thailand}
\author{Walter R. L. Lambrecht}
\email{walter.lambrecht@case.edu}
\affiliation{Department of Physics, Case Western Reserve University, 10900 Euclid Avenue, Cleveland, Ohio 44106-7079, USA}

\begin{abstract}
  Doping of the ultrawide band gap semicodnctor LiGaO$_2$ ($E_g=5.6$ eV) with
  N$_2$, NO and O$_2$ molecules placed in either Ga or Li-vacancies is
  studied using first-principles calculations. These molecular dopants
  are considered as potential acceptors for $p$-type doping. Their optimal
  placement and orientation relative to the lattice is studied as well as their transition levels and energy of formation.  Unfortunately, they are
  found to have deep acceptor level transition states.  We discuss the relation of the transition levels to the one-electron levels, their spin state
  and defect wave functions. They are found to be closely related to those
  of the corresponding vacancies. 
\end{abstract}

\maketitle
\section{Introduction}
LiGaO$_2$ was recently proposed to be a promising  ultrawide band gap
semiconductor. Its band gap is 5.6 eV as verified by optical absorption
measurements \cite{Wolan98,Johnson2011,Chen14} and QS$GW$ (quasiparticle self-consistent $GW$) calculatons.\cite{Radha21ligao2,Boonchun11SPIE}
Similar to $\beta$-Ga$_2$O$_3$ density functional theory (DFT)
calculations  calculations using  a hybrid functional
HSE06 predict that this material could be $n$-type doped by Si or Ge, which 
form shallow donors.\cite{Boonchun19,Dabsamut20} On the other hand, comparing to $\beta$-Ga$_2$O$_3$,
it has a larger gap, somewhat lower valence band masses and a simpler wurtzite related crystal structure, the $Pna2_1$ structure,\cite{Marezio65} in
which all atoms
are tetraehedrally coordinated. It can be grown in bulk form.\cite{Ishii98}

In the only study thus far of the doping opportunities,\cite{Dabsamut20}
$p$-type doping was still found to be problematic, as is also the case
for its parent compound ZnO and for many other oxides, including $\beta$-Ga$_2$O$_3$.
Standard candidate acceptor dopants like N$_\mathrm{O}$ were found to give
too deep levels in the gap and in fact in this case amphoteric character.
Zn in LiGaO$_2$
on the other hand suffers from site competition between the acceptor
Zn$_\mathrm{Ga}$ and donor Zn$_\mathrm{Li}$.  Thus, more involved schemes
need to be pursued for establishing $p$-type doping.  Here we take some inspiration from previous work in ZnO.
\cite{Boonchun13} In ZnO, it is known that a shallow acceptor
exists\cite{Zeuner2002}
that is related to N but is not the simple substitutional defect.
Several  candidates have been suggested. Lautenschlaeger $et$ $al.$\cite{Lautenschlaeger11} proposed that H is involved in a
$\mathrm{N}_\mathrm{O}-\mathrm{H}-\mathrm{N}_\mathrm{O}$ complex. Bang \etal \cite{Bang15} proposed doping with NH$_3$ and a NH$_3$ on the Zn-site. 
Boonchun and Lambrecht proposed a N$_2$ molecule on a Zn-site,\cite{Boonchun13}. Another model is a N$_\mathrm{O}-V_\mathrm{Zn}$ complex.\cite{Liu2012}

The latter was shown to be possible to incorporate making use of the
surface chemistry of adsorption of N on a ZnO O-terminated Zn-polar surface.
In that case, N incorporates preferentially on the Zn sites but then
leads to a O-vacancy in the next layer. After this N$_\mathrm{Zn}-V_\mathrm{O}$
complex is burried inside the sample, the N can easily overcome a barrier
and interchange places  and hop into the adjacent $V_\mathrm{O}$ thereby
creating the desired complex, which was found to be shallow.  This scheme
was made to work in practice by a careful sequence of
growth and annealing treatments by Reynolds {\sl et al.}\cite{Reynolds13,Reynolds14}
and is perhaps one of the most promising routes to $p$-type doping in ZnO.

The alternative proposed scheme with N$_2$ has not yet been pursued experimentally. It is based on the idea of aligning the molecular levels of N$_2$
with those in ZnO. A N$_2$ molecule has 10 valence electrons (2$s$2$p$) in its
neutral charge state and is extremely stable because all the $\sigma$ and $\pi$ bonding states are filled. When placed in a Zn site, it needs to act as a $2+$
ion and thus compared with Zn  including the filled $3d$-band which lies at
about the same energy as the deeper N$_2$ molecule $\sigma$-bonds,
it lacks 2 electrons compared to Zn$^{2+}$. This implies it is a double acceptor. In the single negative charge state of the defect,
(N$_2$)$_\mathrm{Zn}^{-1}$ state it would correspond to
a N$_2^{+1}$ molecule. This state with an unpaired spin has been observed
clearly in Electron Paramagnetic Resonance (EPR) \cite{Garces03}
although these authors assumed it was located on a O-site. However,
Boonchun and Lambrecht\cite{Boonchun13} showed that on the O-site it
would behave as a donor and demonstrated that the Zn-site has qualitatively
the correct $g$-tensor and hyperfine splitting.  While they
proposed it is a shallow acceptor and thus suitable for $p$-type doping,
that depends somewhat on the choice of functional and later calculations
by  Petretto and Bruneval\cite{Bruneval14} found a somewhat different
location of the N$_2$ molecule more strongly bonded to the surrounding O
of the $V_\mathrm{Zn}$  which behaved as a deep acceptor. An experimental
study of the recharging behavior of the N$_2$ in ZnO also proposed that
it is a deep center.\cite{Phillips14}
Nonetheless, the verdict on the shallow or deep
nature of this defect center is  still out and there are alternative explanations for the observed behavior. For example, there might be both
a deep and a shallow configuration of the defect  depending on how
strongly the N$_2$  bonds to the O surrounding it.  Rotational entropy
might favor the less strongly bound form at higher temperatures and
the charging/recharging behavior of the defect taking into account
the higher charge states of the defect has not yet been studied
in detail by computational means.

Here we extend the idea of N$_2$ incorporation on a cation site to  LiGaO$_2$.
There are some differences with the corresponding system in ZnO.
In LiGaO$_2$, the N$_2$ on a Ga site would be a triple acceptor instead
of a double acceptor. The Ga-$3d$ level lies much deeper than the Zn-$3d$
so the alignment with the N$_2$ molecular levels might be different.
The $V_\mathrm{Ga}$ is a defect of high energy of formation, so it might
be less easy to obtain N$_2$ in this site, but the $V_\mathrm{Li}$ has
low energy of formation.  The alignment of the molecular levels
with the host levels is not {\sl a-priori} clear and is investigated here.
Here we study
N$_2$, NO and O$_2$ as molecular doping species  on both cation vacancy
sites and evaluate their formation energy, structural optimization
for different orientations of the molecule and their transition levels.

\section{Computational Methods}
Our study is based on density functional calculations (DFT) using the Heyd-Scuseria-Ernzerhof (HSE) hybrid functional\cite{HSE03,HSE06}
with the standard fraction of exact screened Hartree-Fock type exchange $\alpha$ = 0.25 and range parameter of $\mu$ = 10 \AA, 
giving a gap of $E_g$ = 5.1 eV slightly smaller than the experimental value.\cite{wolan1998chemical,johnson2011electronic, ohkubo2002heteroepitaxial,chen2014growth} 
Since this work extends our previous published results,
we keep the computational approach used here the same as in [\onlinecite{Boonchun19}] and [\onlinecite{Dabsamut20}]. 
The calculations are performed within the Vienna Ab-Initio Simulation Package (VASP).\cite{VASP,KresseVasp1}
The electron ion interactions are described by means of the Projector Augmented Wave (PAW) method.\cite{Blochl94,KresseVasp3}
We use a well-converged energy cutoff of 500 eV for the projector augmented plane waves. 
The dopants are modeled in a 128-atom supercell, for which a shifted single
k-point Brillouin zone sampling was found to be sufficiently converged. 

The analysis of the defect levels follows the standard defect approach as outlined in \eg Freysoldt \etal\cite{Freysoldt14}
The energy of formation of the defect $D^q$ in charge state $q$ is given by
\begin{equation}
	\begin{split}
		E_f(D^q)=E_{tot}(C:D^q)-E_{tot}(C)-\sum_i\Delta n_i \mu_i \\
		+q(\epsilon_v+\epsilon_F+V_{align})+E_{cor}
	\end{split}
\end{equation}
where $E_{tot}(C:D^q)$ is the total energy of the supercell containing the
defect and $E_{tot}(C)$ is the total energy of the perfect crystal supercell.
 The energy for adding or removing atoms from the crystal to a reservoir in the process of producing the defect is represented by the chemical potentials $\mu_i$. The chemical potentials of the host elements that we chose are presented in our previous paper.\cite{Boonchun19} 
The $\Delta n_i$ value represents the change in the number of atoms in the species $i$.  
For the impurities, the chemical potential of N is calculated from the total energy per atom of an isolated N$_2$
molecule, where the chemical potential of O is fixed which respect to the extreme condition as reported in the previous study.\cite{Boonchun19} 
The alignment and image charge corrections
are calculated using the Freysoldt approach.\cite{Freysoldt09}
These energies of formation are 
used to study the transition levels between different
charge states, each with fully relaxed structures but kept at the same volume.

The formation energies of $\rm (N_2)_{Ga}$, $\rm (O_2)_{Ga}$ and $\rm (NO)_{Ga}$ are plotted as function of Fermi level $\varepsilon_F$ in Fig.\ref{fgr:eform} for chemical potential conditions E and F. Note that  chemical potential conditions E and F (see figure 2 in Ref.  \onlinecite{Boonchun19}) correspond respectively to more Li-rich and more Ga-rich conditions but restricted by the formation of competing binary compounds.  Realistic conditions on the O chemical potential corresponding to the typical growth temperature are used in both cases. 
The absolute chemical potentials at point E are $\mu_O=-8.65$ eV,  $\mu_{Li}=-4.01$ eV and $\mu_{Ga}=-6.58$ eV, while the absolute chemical potentials at point F are $\mu_O=-8.65$ eV,  $\mu_{Li}=-6.50$ eV and $\mu_{Ga}=-4.10$ eV. 

The absolute chemical potentials used for the dopants are $\mu_N=-10.24$ eV. However, the formation energies of all three impurities are high with this choice of chemical potentials E and F, implying low solubility in LiGaO$_2$. 
\section{Results}
\subsection{Structural models and relaxation.}
For each molecule we study various configurations of the relative orientation of
the molecule relative to the crystal and fully relax them.
For the N$_2$ and O$_2$ cases, we have first positioned N$_2$ and O$_2$ along the original bond of Ga and O. Both of $\rm (N_2)_{Ga}$ and $\rm (O_2)_{Ga}$, they can be categorized into planar configurations (P1, P2 and P3) and vertical configuration (V1) as shown in Fig.S1 in supplementary information (SI). Similarly to the NO case, the initial position of NO was set along the original Ga-O bond. In this case, we have 6 initial planar structures (planar (P1-a, P1-b, P2-a, P2-b, P3-a, and P3-b) and 2 vertical structures (V1-a and V1-b) as shown in Fig.S2 in SI. The reason why there are more structural models is that
for the $\rm (NO)_{Ga}$ case, we need to distinguish 
how the dipole of the the NO molecule is directed.

\begin{figure}[h]
	\centering
	\includegraphics[width= 9 cm]{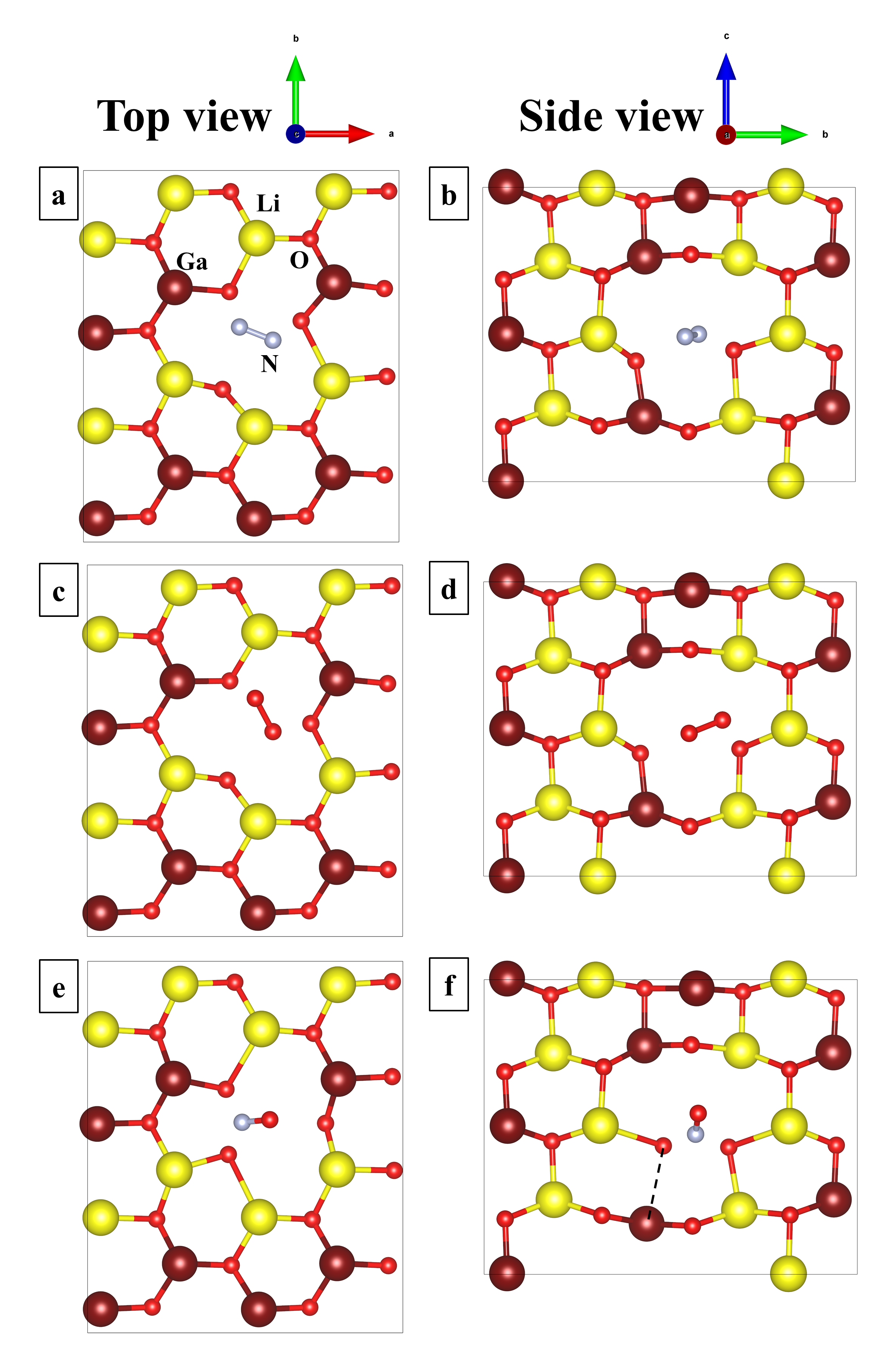}
	\caption{The optimized structure of (a) top view of $\rm (N_2)_{Ga}$, (b) side view of $\rm(N_2)_{Ga}$, (c) top view of $\rm (O_2)_{Ga}$, (d) side view of $\rm (O_2)_{Ga}$, (e) top view of $\rm (NO)_{Ga}$ and (f) side view of $\rm (NO)_{Ga}$. We note that we removed other layers of LiGaO$_2$ to make it easier to spot the molecules. Yellow, brown and red balls represent Li, Ga and O atoms, respectively.}
	\label{fgr:n2ando2}
\end{figure}

Next, we found that the most energetically favourable is the P3 arrangement in the case of N$_2$ and O$_2$. The optimized structure of $\rm (N_2)_{Ga}$ and $\rm (O_2)_{Ga}$ are presented in Fig.\ref{fgr:n2ando2}(a)-(d).
In the top view, we can see only a slight shift of the positions, but in
the side view we may notice a larger shift of the N$_2$ or O$_2$
molecule from their
initial positions. 
For the NO-case, the most energetically favorable is the P2-a arrangement.
The optimized structure is presented in Fig.\ref{fgr:n2ando2}(e)-(f).
Compared to the N$_2$ and O$_2$ cases, we here see a larger distortion of
the O atoms surrounding the vacancy both in the top and side views.
One of the O below has significantly moved closer to the N atom of the NO
to the expense of weakening its bond with the next nearest neighbor Ga.
\begin{figure}[h]
	\centering
	\includegraphics[width= 9 cm]{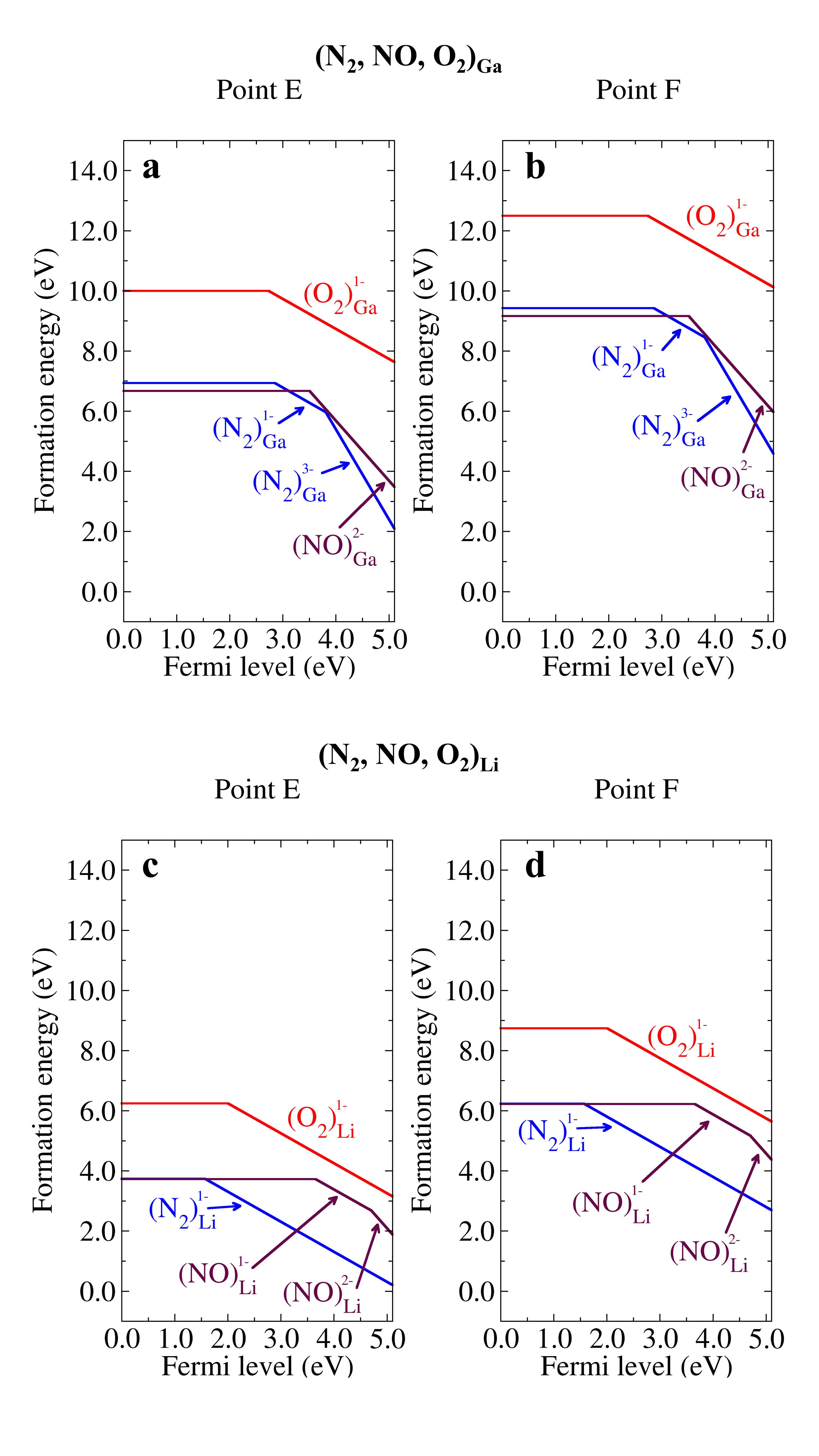}
	\caption{The formation energies of N$_2$, NO and O$_2$ in (a) Ga site for point E, (b) Ga site for point F, (c) Li site for point E and (d) Li site for point F.}
	\label{fgr:eform}
\end{figure}

\subsection{Energies of formation and transition levels}
The energies of formation for different charge states, choosing the
configuration with lowest structural energy  are given in Fig. \ref{fgr:eform}
for two cases of chemical potentials, more Li-rich (E) and more Ga-rich (F) cases. Both points are chosen to pick
a realistic O chemical potential corresponding to the growth conditions
as explained in Ref. 6.

They show first that the energies of formation on the Ga site (Fig.\ref{fgr:eform}(a)-(b)) are relatively high. The reason for this is simply that the Ga vacancy itself is a high-energy of formation defect as reported in our previous study.\cite{Boonchun19} One could in principle still realize these defect
cases by first creating vacancies using for example high-energy irradiation and
subsequently introduce the dopants by reactions from the surface inwards,
by exposure to N$_2$, NO or O$_2$ gas.

Second, they show that N$_2$ occurs in three charge states depending
on the position of the Fermi level, neutral, $-1$ and $-3$.  This indicates
a negative $U$ behavior between $-1$ and $-3$ charge states. Interestingly,
in the corresponding Ga-vacancy, the $-2$ charge state also has  already
a narrow energy range of stability. 
Likewise, NO occurs in neutral and $-2$ charge states only.   However, 
O$_2$ occurs only in neutral and $-1$ charge states and exhibits
no negative-$U$ behavior.

Third, the transition levels all are quite deep with the lowest $0/-1$
transition level occurring at about 2.85 eV above the valence band maximum (VBM) for the N$_2$ case. 

The energy of formation
results for the various molecules on the Li site are given
in Fig.\ref{fgr:eform}(c)-(d). These have slightly lower energies of formation consistent with the
lower energy of formation of the $V_\mathrm{Li}$ as reported in our previous study.\cite{Boonchun19} Structurally, there
is ample room for these molecules in both types of vacancy sites.
The transition levels are slightly shallower for the Li site occupation.
For example, the $0/-1$ transition level for (N$_2$)$_\mathrm{Li}$
is 1.57 eV instead of (N$_2$)$_\mathrm{Ga}$ where it is 2.85 eV.
The (N$_2$)$_\mathrm{Li}$  exists only in 0 and $-1$ states and
the same holds for the O$_2$ case, while the NO molecule exists in
three charge states, 0, $-1$, $-2$. 

\subsection{One-electron levels and defect wave functions.}
\begin{figure}
  \includegraphics[width=9cm]{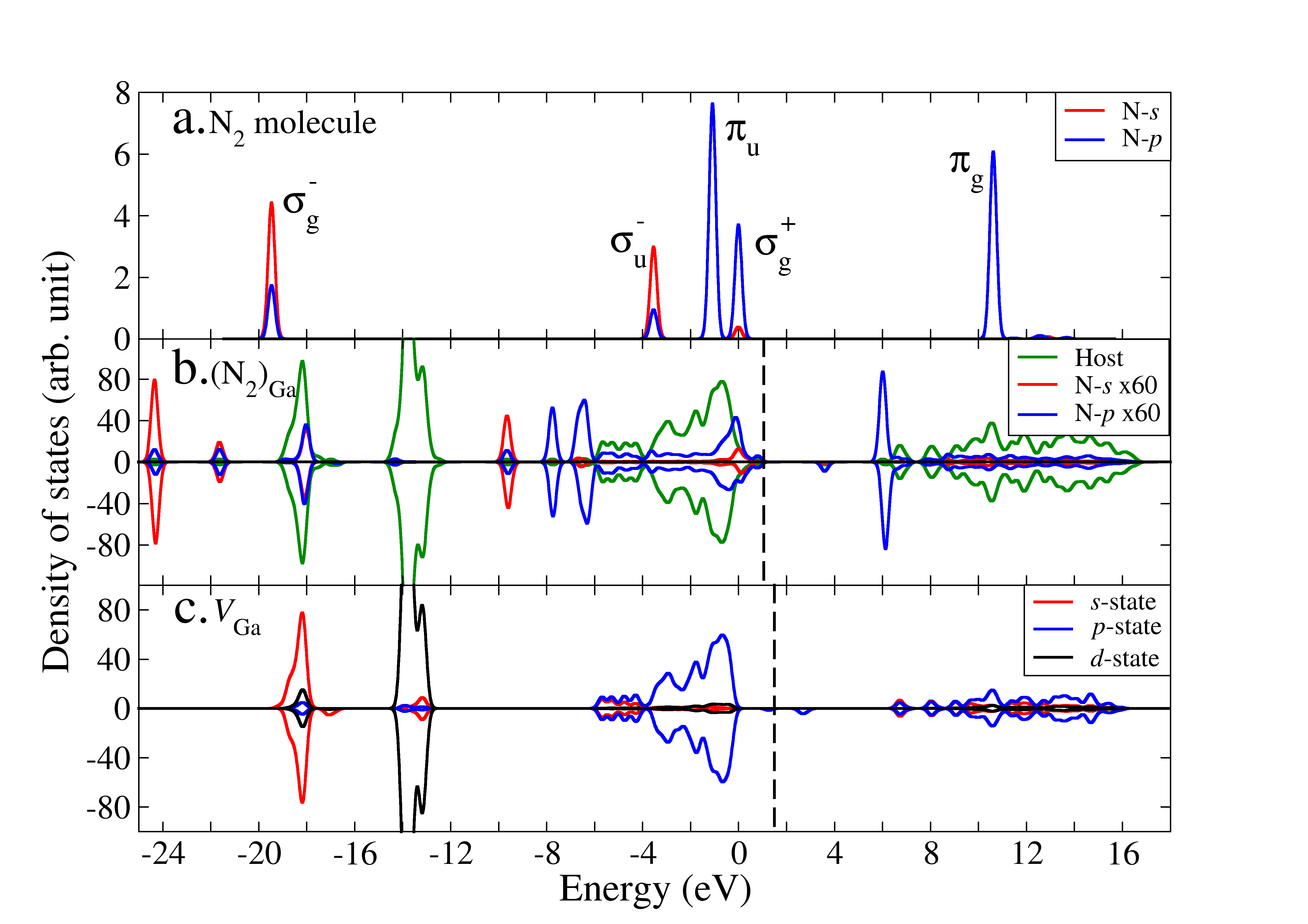}
  \caption{PDOS of N$_2$ in the Ga-vacancy in the neutral charge state of the defect:  (a) isolated N$_2$,
    (b) interacting system, and (c) Ga-vacancy. We note that the dash lines indicate the highest occupied state.
    \label{fgr:dosn2ga}}
\end{figure}

\begin{figure}
	\includegraphics[width=5cm]{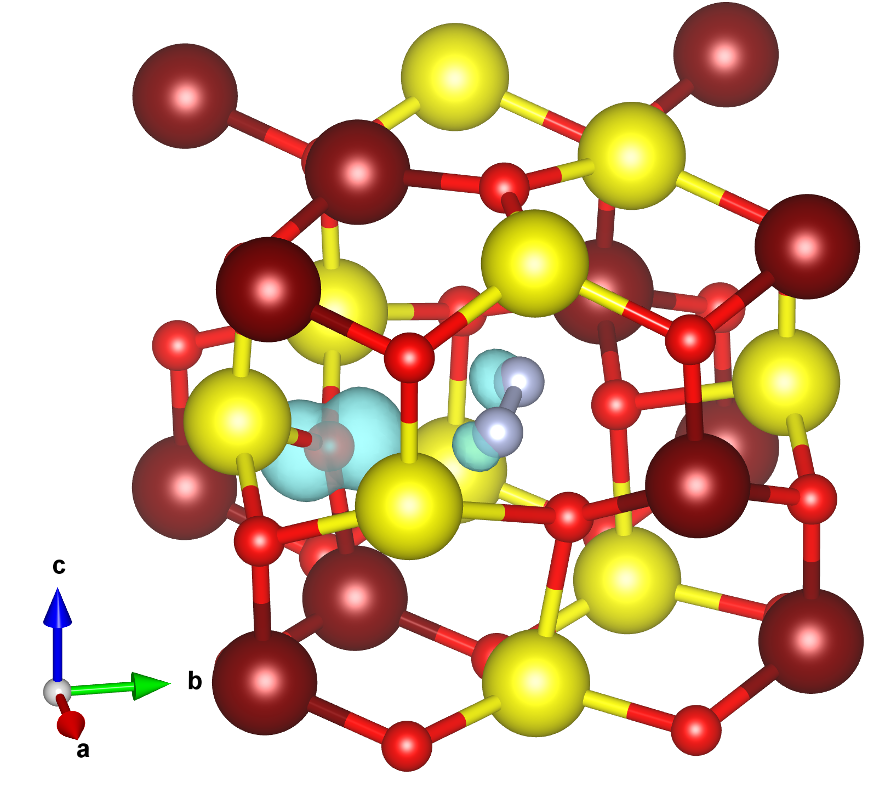}
  \caption{Net spin density of  (N$_2$)$_\mathrm{Ga}$ in the neutral
    charge state.  \label{fgr:spinn2ga}}
\end{figure}

In this section, we examine the one-electron levels, their defect
wave functions, and partial densities of states (PDOS) to clarify
the spin states of the defects and their relation to the transition levels.

In Fig. \ref{fgr:dosn2ga} we show the PDOS  of N$_2$ placed in the
Ga-vacancy in panel (b), while panels (a) and (c) show the isolated
molecule and the Ga-vacancy PDOS respectively.
In panel (a), we have labeled the peaks according to the standared nomenclature
for molecular eigenstates. 
First, we recognize the lowest empty molecular orbital (LUMO)  $\pi_g$ antibonding state  lying above 10 eV. Next, the $\sigma_g^+$ state forms
the highest occupied molecular orbital (HOMO). We see that it 
consists mostly of N-$2p$  orbitals with  a small N-$2s$ contribution. The HOMO-1 level is the $\pi_u$ bonding state, which has pure $p$ character and
is doubly degenerate and thus has twice the peak height of the HOMO.
The next level is the $\sigma_u^-$ state which
is a bonding state between N-$2s$ and N-$2p_z$. Its antibonding counter part
lies at higher energy above the CBM and is not shown. 
The $\sigma_g^-$ state at $-20$ eV is the bonding counterpart to the
$\sigma_g^+$ state and has mostly N-$2s$ character with a little bit of N-$2p$.
Here, the labels $g,u$ refer to 
even or odd with respect to inversion and $\pm$ refers to (anti)bonding
character. We need this extra label only for $\sigma$ states because there
is more than one $\sigma_g$ and $\sigma_u$. 
From panel (b) we can see that there is a net spin-polarization
in the neutral charge state, which is related to its odd number of electrons.
We can still clearly recognize the $\pi_g$ antibonding state of the N$_2$ molecule, lying at the conduction band minimum (CBM). A minority spin state
is seen at mid gap and states of both majority and minority spin character
occur just above the VBM.  We here aligned the deep O-$2s$ states with
those of the host system containing the Ga-vacancy shown in panel (c)
and indicate the Fermi level at zero temperature (highest occupied state)
by the dashed line. We can see several defect levels of minority spin
also in panel (c) for the neutral Ga-vacancy with only the lowest one of these occupied. These levels look
dangling bond like (see Fig.S6 in SI) but are spread over several
neighboring O sites. 

The N-$2p$ projected
PDOS of the N$_2$ molecule inside the Ga-vacancy shows a peak about 1 eV
below the VBM with a net difference between spin-up and spin down,
indicating that part of the spin  density must be localized on the molecule.
However, the bonding $\pi_u$ state of the molecule are more perturbed
by the interaction with the host. 
It appears  that this state forms both bonding and antibonding states with
the O-$2p$ valence states with the bonding states lying just below the O-$2p$
bands and the antibonding ones slightly below the VBM. In fact, we find
three peaks localized on the molecule lying just below the O-$2p$ valence band,
two with predominantly N-$2p$ character and one with predominantly
N-$2s$ character. The $\pi_u$ state in the molecule has pure $p$ character
but when placed in the Ga-vacancy it  acquires a small amount of $s$ character
and the peak height is now the same as for the $\sigma_g^+$ state because
its weight is also partially found in the states just below the VBM and
distributed to some extent throughout the valence band. 
The states near $-14$ eV are seen to be the Ga-$3d$ states
and some hybridization between the deep lying $\sigma_g^-$ states with
the O-$2s$ band can  also be seen.  All of this indicates a significant
interaction between the molecular states and the vacancy states.
The net spin density in the neutral charge state of (N$_2$)$_\mathrm{Ga}$
is shown in Fig. \ref{fgr:spinn2ga} and confirms this picture.

\begin{figure}[h]
	\centering
	\includegraphics[width= 8.5 cm]{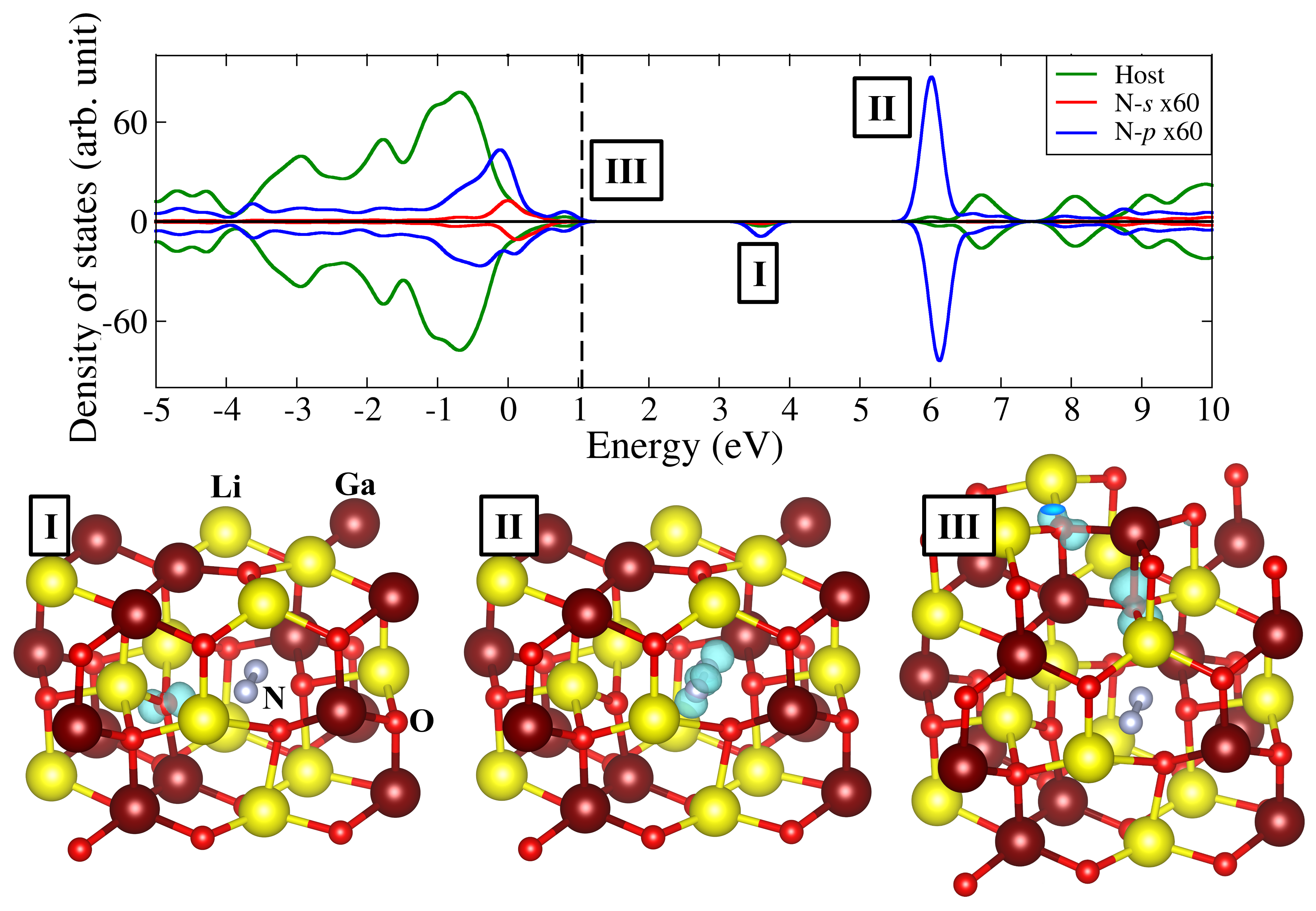}
	\caption{The partial density of state (PDOS) and the partial charge density of (N$_2$)$_\mathrm{Ga}$. We note that the dash line separates the occupied and unoccupied state.}
	\label{fgr:dosandparchg_n2}
\end{figure}

In Fig.\ref{fgr:dosandparchg_n2}, we show the partial densities of states (PDOS)
and defect wavefunctions modulo squared for several energy ranges
corresponding to defect levels for the N$_2$ on Ga site.
We note several defect levels or resonances in the bands introduced by
the molecule insertion into the vacancy.  First, the state (labeled II)
corresponds clearly to the molecular $\pi^*$ antibonding state and lies
just below the CBM of the host crystal.
A resonance also occurs just below the VBM
and corresponds to the highest occupied molecular orbital (HOMO)
of the N$_2$ molecule and has $\sigma$-bonding character.
In addition to these molecular levels, we recognize a defect level
(labeled I) which is localized on one of the O surrounding the defects.
It is accompanied by a  polaronic distortion localizing the wave function
on only one of the O nearest neighbors of the Ga vacancy in which
the N$_2$ is inserted.  A defect level closer to the VBM, (labeled III)
still shows also localization on one of the nearest neighbor oxygens
but spreads somewhat further to further neighbor O.
All of these were obtained for the neutral defect.

For the N$_2$ on the Ga site, which has 3 holes in the defect or molecule levels
compared to the perfect crystal since it can exist  also in
-1, -2 and -3 charge states by successively adding electrons, we have
also examined the possibility of a higher spin $S=3/2$ state with three
parallel spins. This configuration, however, was found to have 1.53  eV higher energy and can thus safely be excluded from further consideration.
We find from the corresponding PDOS that the N-$p$ related
states occur deeper below the VBM, as shown in Fig.S6 in SI.

\begin{figure}[h]
  \includegraphics[width=9cm]{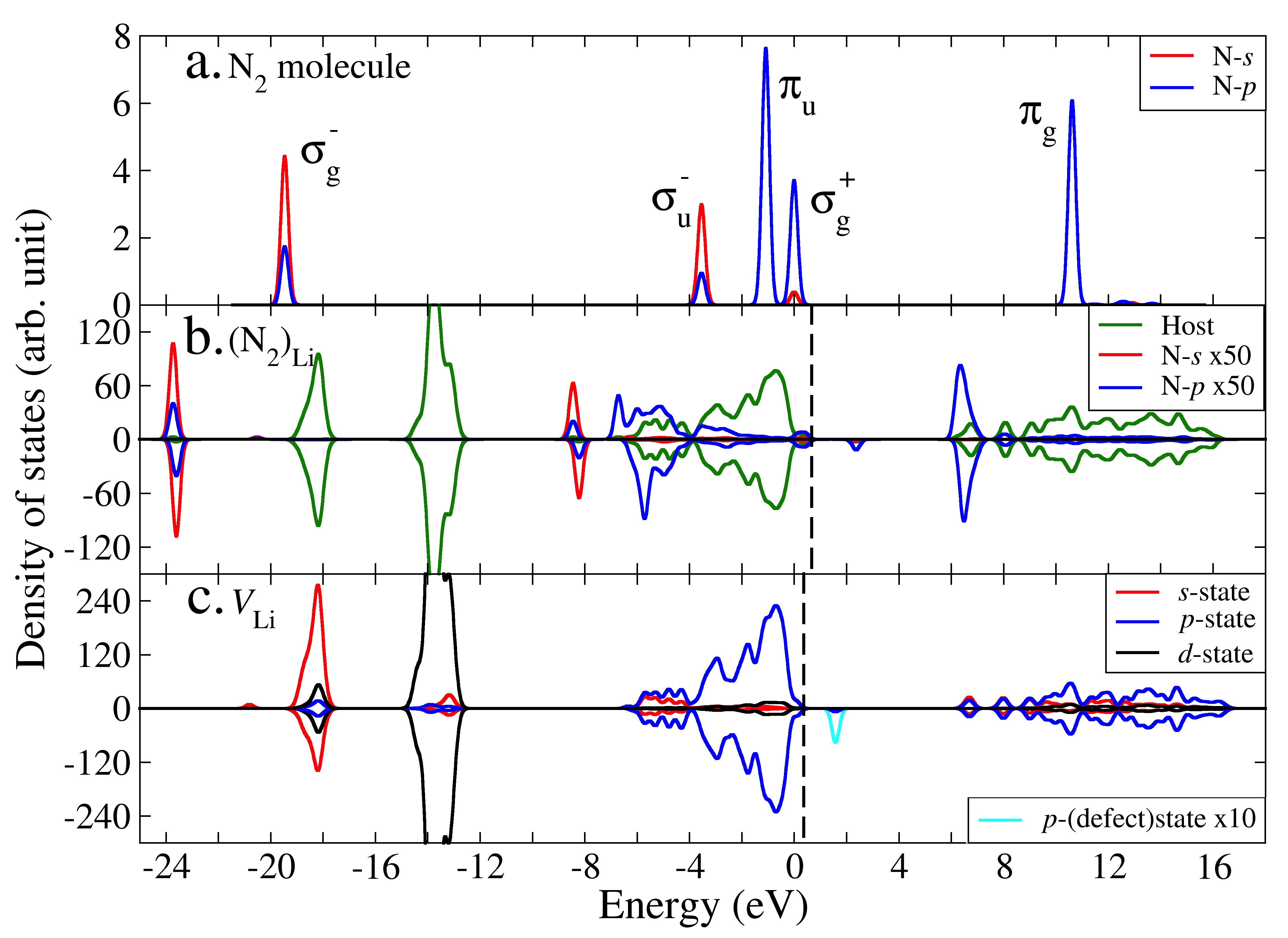}
  \caption{PDOS of N$_2$ on Li-site in neutral charge state. (a)
    isolated molecule, (b) molecule in the Li-site and (c) Li-vacancy. We remark that the dash lines separate the occupied and unoccupied states.
    \label{fgr:n2li}}
\end{figure} 

\begin{figure}
	\includegraphics[width=5cm]{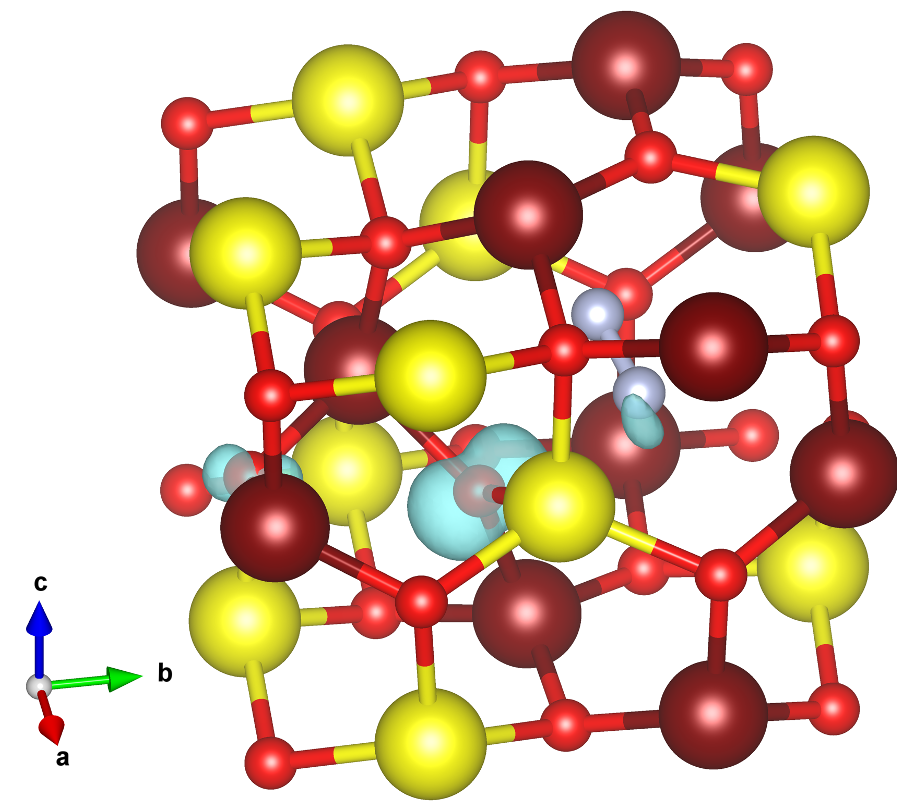}
  \caption{Net spin density for (N$_2$)$_\mathrm{Li}$ in the neutral charge state.
  	 \label{fgr:spinn2li}}
\end{figure}

For N$_2$ placed  on the Li-site we similarly show the  comparison
between the N$_2$ PDOS in the isolated molecule  and in the Li site
and with the Li-vacancy case, again, all for the neutral charge state.
The $\pi_g$ state is again most readily identified as lying near the
CBM  at about 6 eV.  Defect states related to the Li-vacancy
are also found in the gap. However, in this case the N$_2$ HOMO
state is located near the bottom of the O-$2p$ states
and much less interaction is observed than for the Ga-vacancy.
This indicates a different electrostatic potential in which the N$_2$
molecule finds itself. In fact, it is a less repulsive potential because
removing a Li$^+$ is less repulsive than removing a Ga$^{3+}$, 
thereby
placing these states deeper relative to the O-$2p$ bands.  The  molecular
states are thus differently aligned with respect to the LiGaO$_2$ host
bands when the N$_2$ is located in the Ga than in the Li vacancy.
A weaker interaction is also observed for the deep $\sigma_g^-$ state
with the O-$2s$ band.   There is is still a net spin-density as
shown in Fig. \ref{fgr:spinn2li}.

Until now we have focused on the neutral charge state and on the N$_2$
molecule related states and their interaction with  their surroundings.
However, it is also clear that several states exist in the gap and
inspection of their wave functions shows that these states are primarily
localized on a single oxygen  and are closely related to the corresponding
vacancy states. Adding an electron in the negative charge states
makes the system non-spin polarized as there is now an even number of electrons.
We checked that the high-spin $S=1$ configuration for the $q=-1$ charge
  state has higher energy by 1.05 eV. Again, we find that the occupied
  one-electron
  states of the N$_2$ molecule occur farther below the VBM.
  We can study the addition of an electron in two steps, first keeping
  the structure frozen in that of the neutral charge state and subsequently
  relaxing the atomic positions.  We find that the levels in the gap
  are spread deeper into the conduction band in the frozen geometry of the neutral state and move closer to the VBM upon relaxation. This is shown in Fig.S8 in the SI.
Likewise in the $-2$ charge state, we have again a  net spin polarization and so on.  After filling the empty states of the molecule, the next electrons
essentially are located in the vacancy related dangling bond states and
correspond closely to the corresponding vacancy case. However for the
$V_\mathrm{Ga}$  we find several states  with localization on more than one
O-neighbor to the vacancy.

\begin{figure}
  \includegraphics[width=9cm]{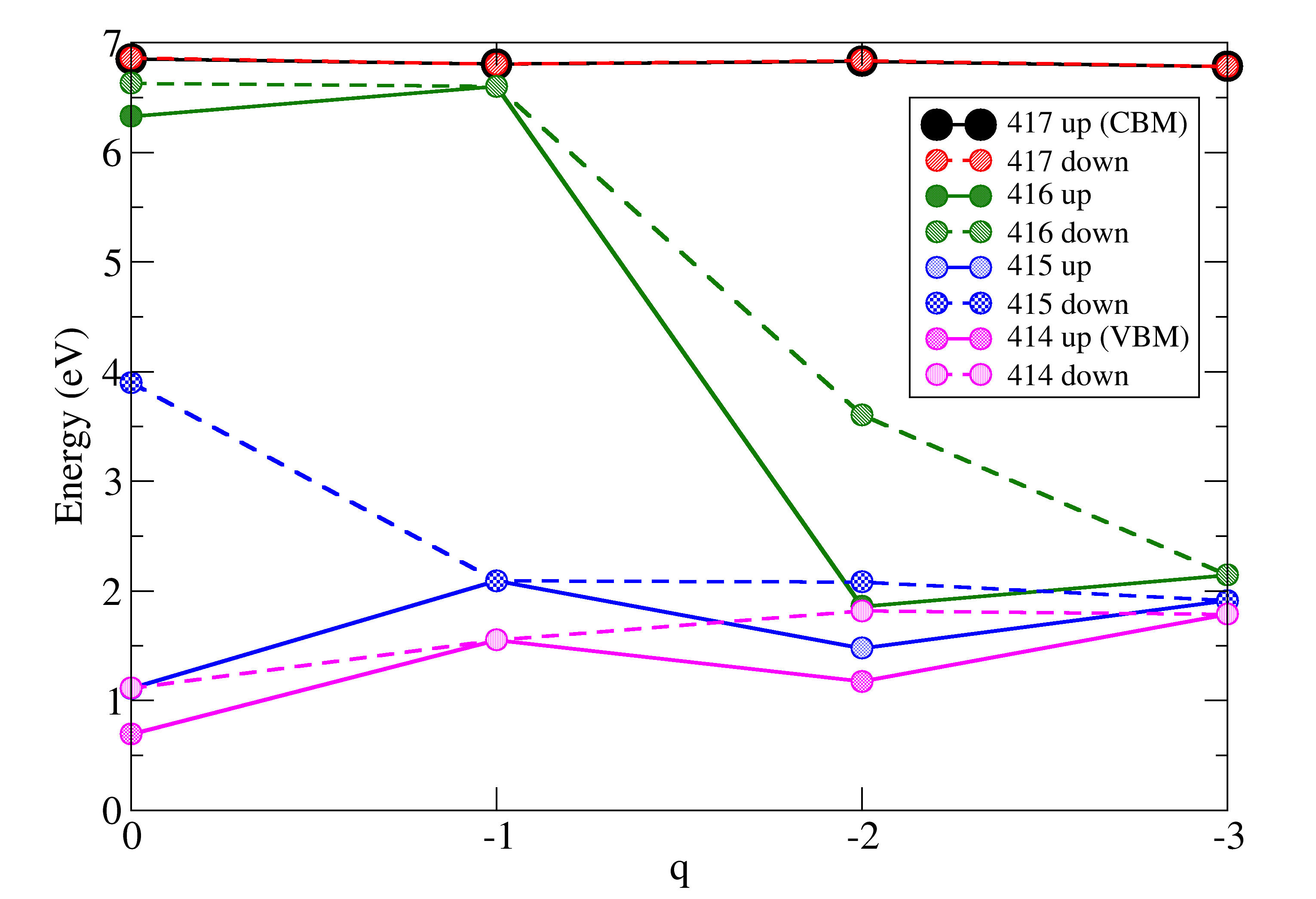}
  \caption{Energy levels in the supercell as function of charge state
    at the shifted ${\bf k}$-point, labeled by the eigenvalue number.
    Solid lines connect spin up states and dashed lines spin down states. \label{levelsq}}
\end{figure}

The one-electron energy levels lying inside the band gap
are shown in Fig.\ref{levelsq} as function of charge state.
They are shown for both spin states. One can see that in the neutral charge state only  one defect level (No. 415 $\uparrow$) close to the VBM (levels 414 for $\uparrow$
and $\downarrow$ spin) is occupied while three remain empty.
In the $q=-1$ charge state, the lower one of these, which is the minority spin
level 415, is lowered significantly while it becomes occupied.  Likewise, in
the $q=-2$ and $q=-3$ charge states, additional levels come down in energy
as they become occupied.  This reflects the fact that in hybrid functional,
due to the partial inclusion of exact exchange, the potential is
orbital-dependent and occupied levels shift down relative to empty levels.

We should here caution the reader that these one-electron levels, 
also seen as peaks in the PDOS,  reflect
the band positions in the 128 cell model, which is not truly reflective
of the dilute limit.  An electrostatic correction $-2q\alpha/\varepsilon L$ applies to
the single particle levels for charged defect states due to their interaction with the image charges and the neutralizing background, similar to the
well-known corrections to the energies of formation.
These corrections allow one to extrapolate to the dilute limit and  would 
shift the one-electron levels of the isolated defect deeper into the
gap with a shift proportional to their charge and inversely proportional
the size of the supercell $L$. Here $\alpha$ is a Madelung constant and
$\varepsilon$ is the dielectric constant of the host. 
Because these defect one-electron levels are pushed  deeper in the gap
than indicated by the PDOS, filling them only occurs when the Fermi level
is also rather deep in the gap and this essentially explains the deep
nature of the transition levels even though rather shallow states
are seen in the PDOS figures or for the occupied one-electron levels. 

We can further examine the PDOS of these different charge states and
the associated wave functions. These details are given in Fig.S9 in the SI
for charge states -1,-2, -3. 
The main conclusion is that the peaks in the PDOS can
be associated with the levels given in Fig. \ref{levelsq} and
their associated states show mostly vacancy type localized states with
in some cases a partial contribution on the molecule and
for states closer to the VBM a wider spread  farther away from the
defect site.
We thus find that the states in the gap are closely related to those of the corresponding vacancy and have generally the character of
dangling bond states on the surrounding oxygen atoms. The relaxation tends to localize them on a single oxygen rather than being distributed over several
oxygens.  This indicates the polaronic nature of these defect states.

Moving on to the NO and O$_2$ as inserted molecules, the changes can basically
be understood in terms of the added valence electron in each step.
The PDOS of the NO and O$_2$ on the Ga and Li site in the neutral charge state  
are shown in SI. They indicate that the NO molecule in the neutral charge
state is similar to the N$_2$ in the $-1$ state and the O$_2$ molecule in the
neutral state is similar to the N$_2$ in the $-2$ state.  The defect levels
in the gap in all these cases are mostly localized on a single oxygen neighbor
of the vacancy in which the molecule is placed.

\section{Discussion}
In the present calculations, we find deep levels and polaronic acceptor behavior
for each of these molecules when inserted in either a Li or Ga vacancy.
This is manifested in the localization of the defect wave functions each
time on a single O. 
In part, this behavior is emphasized by the hybrid
functional choice. All of these results were obtained with the
standard HSE setting of 25 \% exact exchange.  As a final check on our
results, we  therefore investigated the satisfaction of the
generalized Koopmans theorem (gKT) by calculating the non-Koopman's energy\cite{ivady2013role}, 
\begin{equation}
	\begin{split}
		E_{NK} = \varepsilon_N - E_I = \varepsilon_N - (E_N - E_{N-1})
	\end{split}
\end{equation}
where $\varepsilon_N$ is a localized state's Kohn-Sham quasiparticle energy in an N electron system, $E_I$ is the system's ionization energy, which is equivalent to the difference between the total energies $E_N$ and $E_{N-1}$ of the N and N-1 electron systems with defect, respectively. 
In the Table \ref{tbl:1}, we calculated the non-Koopman's energy ($E_{NK}$) for 20\%, 25\% and 30\% HF mixing parameters from our previous Sn$_\mathrm{Ga}$ calculation.\cite{Dabsamut20} In the case of the gKT, $E_{NK}$ should equal be zero. As a result, we found that $E_{NK}$ for 30\% HF mixing parameter is almost zero (-0.012 eV), indicating that gKT is fulfilled. Moreover, the energy gap from this calculation is very close to the experiment value as presented in Table \ref{tbl:1}. Although this indicates
that a 30 \% HF would have been preferable, we have kept the 25 \% results
to be consistent with our previous papers. 
Even for the 25\% HF mixing parameter, $E_{NK}$ is already relatively small (-0.121 eV), thus the self-interaction error may be considered small
in all the present calculations. 

\begin{table}[h]
	\centering
	\caption{The calculated energy gap ($E_g$), the localized state's Kohn-Sham quasiparticle energy ($\varepsilon_N$), the system's ionization energy ($E_I$) and the non-Koopman's energy ($E_{NK}$) for 20\%, 25\% and 30\% HF mixing parameters in Sn$_\mathrm{Ga}$ system.}
	\label{tbl:1}
	\begin{tabular*}{0.48\textwidth}{@{\extracolsep{\fill}}cccccc}
		\hline 	\hline
	    & &	\multicolumn{3}{c}{HF mixing parameter} & \\
		& & 20\% & 25\% & 30\% & Exp. value \\
		\hline
		& $E_g$ (eV) & 4.70 & 5.10 & 5.50 & 5.3-5.6\cite{wolan1998chemical,johnson2011electronic, ohkubo2002heteroepitaxial, chen2014growth} \\
		& $\varepsilon_N$ (eV) & 3.818 & 4.010 & 4.070 & \\
		& $E_I$ (eV) & 4.057 & 4.132 & 4.082 &  \\
		& $E_{NK}$ (eV) & -0.240 & -0.121 & -0.012 &  \\

		\hline 	\hline
	\end{tabular*}
\end{table}

\section{Conclusions}
In this paper we have examined the behavior of inserted diatomic molecules N$_2$, NO and O$_2$ in both the Li and Ga site.  We found that these systems
have high energy of formation, especially for the Ga site, which is
related to the corresponding vacancy formation energy.
Even though these systems all behave as acceptor systems, we found deep
transition levels in the gap, which are not compatible with $p$-type doping.
Analysis of the PDOS and one-electron levels shows that as successive levels become occupied in the negative charge states, they move closer to the  VBM.
This is a direct result of the hybrid functional used which lowers
occupied {\sl vs. } empty levels. 
While the occupied ones  appear much shallower than the  transition levels,
this is an artifact of the supercell size. Corrections for the one-electron
levels in charged systems will tend to move these levels deeper in the gap
proportional to their charge state, so that the transition levels and
one-electron levels become consistent with each other in labeling these systems as deep acceptors. The behavior of NO and O$_2$  in charge state $q$
correspond to the corresponding $q-1$, $q-2$ states of N$_2$. We find a
stronger interaction of the N$_2$ molecular states with the O-states of the host for the Ga than for the Li position  in terms of the occupied states
of the molecule. The defect levels in the gap are closely related to the
corresponding vacancy levels and are dangling bond like and characterized
by polaronic distortions localizing the defect wave functions on a single
O neighbor to the vacancy.  We find that the Koopmans theorem is
already fairly well satisfied by the 25 \% exact exchange hybrid functional
used here and in our previous work on native defects, 
although  in future work a 30 \% model would both improve the band gap
and the satisfaction of Koopmans theorem.

\acknowledgements{The work at CWRU was supported by
  the U.S. National Science Foundation under grant No. 1755479. 
K.D. was supported by the National Research Council of Thailand (NRCT), No. NRCT5-RGJ63002-028. A.B. has been funded by the Office of National Higher Education Science Research and Innovation Policy Council (NXPO), Thailand, through Program Management Unit for Competitiveness (PMU C), Contract Number C10F630073. Calculations were in part performed at the Ohio Supercomputer Center.}
  
\bibliography{arXiv}
 \end{document}